\documentclass[aps,twocolumn,groupedaddress,showpacs,floats]{revtex4}
\usepackage[dvips]{graphicx}
\usepackage{bm}
\begin{document}

\title{Scale Separation Scheme for Simulating Superfluid Turbulence: Kelvin wave Cascade}

\author{Evgeny Kozik$^{1}$ and Boris Svistunov$^{1,2,3}$}
 \affiliation{
${^1}$ Department of Physics, University of
Massachusetts, Amherst, MA 01003 \\
${^2}$ Russian Research Center ``Kurchatov Institute'', 123182
Moscow, Russia\\
${^3}$ Kavli Institute for Theoretical Physics, University of
California, Santa Barbara, CA 93106}


\begin{abstract}
A Kolmogorov-type cascade of Kelvin waves---the distortion waves
on vortex lines---plays a key part in the relaxation of superfluid
turbulence at low temperatures. We propose an efficient numeric
scheme for simulating the Kelvin wave cascade on a single vortex
line. The idea is likely to be generalizable for a full-scale
simulation of different regimes of superfluid turbulence.  With
the new scheme, we are able to unambiguously resolve the cascade
spectrum exponent, and thus to settle the controversy between
recent simulations \cite{Vinen_2003} and recently developed
analytic theory \cite{SvK_2004}.
\end{abstract}





\maketitle

The superfluid turbulence (for an introduction see \cite{Donnelly,
Cambridge}) can be viewed as a tangle of quantized vortex lines.
In the case of an essentially finite temperature, the superfluid
turbulence is perfectly described by the local induction
approximation (LIA), as it has been shown in pioneering
simulations by Schwarz \cite{Schwarz_85, Schwarz_88}. LIA reduces
the problem of finding the velocity of an element of the vortex
line to its local differential characteristics:
\begin{equation}
\dot{\textbf{s}} = \beta \, \textbf{s}' \times \textbf{s}'' +
\mbox{ friction} \; , \label{LIA}
\end{equation}
\begin{equation}
\beta =  {\kappa \over 4 \pi} \, \ln (R/a_0)\; , \label{beta}
\end{equation}
where $\textbf{s}\equiv \textbf{s}(\xi,t)$ is the evolving in time
radius-vector of the vortex line element, the parameter $\xi$
being the arc length, the dot and prime denote differentiation
with respect to time and $\xi$, respectively; $\kappa$ is the
quantum of velocity circulation; $R$ is the typical curvature
radius (which is of the same order that the typical interline
spacing), and $a_0$ is the vortex core radius. A necessary
condition of applicability of LIA is
\begin{equation}
\ln (R/a_0) \gg 1 \; . \label{ln}
\end{equation}

At zero temperature, a crucial role in the dynamics of superfluid
turbulence is played by the Kelvin wave cascade
\cite{Sv_95,Vinen2000,Davis,Kivotides,Vinen_2003,SvK_2004}. LIA
fails to properly describe a considerable part of the cascade
inertial range (see, e.g., \cite{SvK_2004} and references
therein). In this case, one has to use the description in terms of
full Biot-Savart equation (BSE):
\begin{equation}
\dot{\textbf{s}} = {\kappa \over 4 \pi} \int (\textbf{s}_0 -
\textbf{s}) \times  {\rm d}\textbf{s}_0 / |\textbf{s}_0 -
\textbf{s}|^3 \; . \label{BS}
\end{equation}
Here the evolving in time radius-vector of the vortex line
element, $\textbf{s}\equiv \textbf{s}(\zeta,t)$, is  labelled by
some parameter $\zeta$ (not necessarily the arc length), the
vector $\textbf{s}_0$ is of the same physical meaning as
$\textbf{s}$, understood as an integration variable; the
integration is over all the vortex lines; ${\rm d}\textbf{s}_0=
(\partial
 \textbf{s}_0 /\partial \zeta)\, {\rm d} \zeta $. The two conditions of applicability of
Eq.~(\ref{BS}) are: (i) the absence of the normal component
implying the absence of the friction term, and (ii)
\begin{equation}
R \gg a_0 \; . \label{cond}
\end{equation}
The need to use the full Biot-Savart law in dealing with the
Kelvin wave cascade had been recognized in
Refs.~\cite{Kivotides,Vinen_2003}. It is worth noting that the
Kelvin wave cascade is not the only problem where non-local vortex
line interactions play a crucial part. The other examples include
the conditions of stability of vortex knots revealed by Ricca {\it
et al}. \cite{Ricca}: the vortex knots that are unstable under LIA
acquire a greatly increased lifetime when BSE in used. In the
simulation by Samuels \cite{Samuels_92}, the non-local interaction
of vortex rings in laminar flow of He II results in overall
parabolic superfluid velocity profile matching the normal-fluid
velocity profile. In the problem of superfluid turbulence decay at
length scales greater than the interline spacing, the non-local
interactions of vortex lines result in Kolmogorov cascade of
eddies mimicking the behavior of the classical turbulence even
without the normal component \cite{Nore, Tsubota_2002,
Barenghi_2002}.

In theoretical studies of turbulent vortex dynamics, an analytic
solution is normally not available, and numeric simulations are of
crucial importance. Clearly, the computational price of a direct
simulation of BSE is macroscopically larger than that of LIA,
since to perform one step in time for a given element of a line
one has to address all the line elements. In this Letter, we
observe that this extensive addressing is actually extremely
excessive and can be easily eliminated at the expense of a {\it
controllable} systematic error. The physical reason for that is as
follows. The interactions at a distance $r$ are significant only
for the Kelvin waves of the wavelength $\lambda \gtrsim r$. The
characteristic time of evolution of a harmonic of the wavelength
$\lambda$ scales like $\lambda^2$, which means that during the
time interval much shorter than $\lambda^2$ the configuration of
these harmonics remains essentially unchanged. Correspondingly,
the maximal reasonable frequency of addressing the distances $\sim
r$ from a given element of the vortex line scales like $1/r^2$.
This leads to a numeric scheme where the long-range interactions
of Kelvin waves do not reduce the simulation efficiency. We employ
this scheme to simulate the Kelvin wave cascade on a single vortex
line at zero temperature. Our data for the cascade spectrum are in
an excellent agreement with the predictions of analytic theory
\cite{SvK_2004}.

{\it Scale separation scheme}.---To be specific, we consider the
zero-temperature dynamics of Kelvin waves on a straight  vortex
line. We separate different scales of distance in the integral in
Eq.~(\ref{BS}) by writing BSE in the form
\begin{equation}
\dot{\textbf{s}} = Q_{1}+Q_{2}+Q_{3}+ \ldots + Q_{n}+ \ldots \; ,
\label{BS1}
\end{equation}
where
\begin{equation}
Q_{n }(\textbf{s}) = {\kappa \over 4 \pi} \int_{2^{n-1}a_0<|{\bf
s}_0 - {\bf s}|<2^{n}a_0} { (\textbf{s}_0 - \textbf{s}) \times
{\rm d}\textbf{s}_0 \over |\textbf{s}_0 - \textbf{s}|^3} \; .
\label{int_n}
\end{equation}
Strictly speaking, Eqs.~(\ref{BS1})-(\ref{int_n}) are meaningful
only for $n \gg 1$, since at distances $\lesssim a_0$ from the
vortex core the velocity field is model-dependent. With the same
accuracy, one may replace in Eq.~(\ref{BS1}) the $n\sim 1$
non-local terms with a local term. (As we do it in our simulation
scheme.---See below.)
\begin{figure}[htb]
\includegraphics[width = 0.99\columnwidth,keepaspectratio=true]{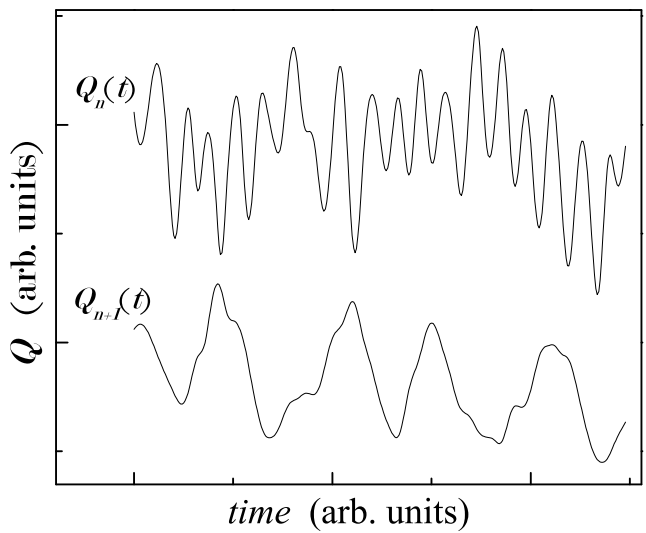}
\caption{Comparing functions  $Q_{n}(t)$  and $Q_{n+1}(t)$ for a
fixed vortex line element. The data are obtained by simulating
Kelvin wave cascade with full BSE. } \label{fig_rhs}
\end{figure}

The idea behind the separation of scales of distance in
Eq.~(\ref{BS1}) is clearly seen in Fig.~1, where we compare two
functions, $Q_{n}(t)$ and $Q_{n+1}(t)$, in the regime of Kelvin
wave cascade. The characteristic period of variation of the
function $Q_{n+1}(t)$ is approximately $4$ times larger than that
of $Q_{n}$. This is precisely what one would expect on the basis
of qualitative analysis, since the typical wavelength of harmonics
contributing to $Q_{n}$ is $\lambda_n = 2^n a_0$, which leads to
the estimate
\begin{equation}
\tau_n \propto  \lambda_n^2 \label{est}
\end{equation}
for the typical time of variation of $Q_{n}$. The contributions to
$Q_{n}$ from the harmonics with $\lambda \ll \lambda_n$  are
suppressed by the denominator in Eq.~(\ref{int_n}) in combination
with a random interference in the numerator.

The behavior of $Q_{n}$'s suggests the following numeric
procedure. Each $Q_{n}(t)$ is approximated by a piecewise
polynomial function $\tilde{Q}_{n}(t)$ based on an appropriate
mesh $\{ t^{(n)}_j\}$, $j=1,2,3,\ldots$, and corresponding
approximate values of $Q_{n}(t=t^{(n)}_j)$ obtained by integrating
BSE with $Q_{n}(\textbf{s},t) \to \tilde{Q}_{n}(\textbf{s},t)$. In
the present study, we take the values of $Q_n$ calculated at the
time moments $t^{(n)}_{j-1}$ and $t^{(n)}_{j}$, and  use the
linear extrapolation to get the approximation $\tilde{Q}_{n}(t)$
for $Q_n(t)$ at any $t$ in the interval
$[t^{(n)}_{j},t^{(n)}_{j+1}]$.

\begin{figure}[htb]
\includegraphics[width = 0.99\columnwidth,keepaspectratio=true]{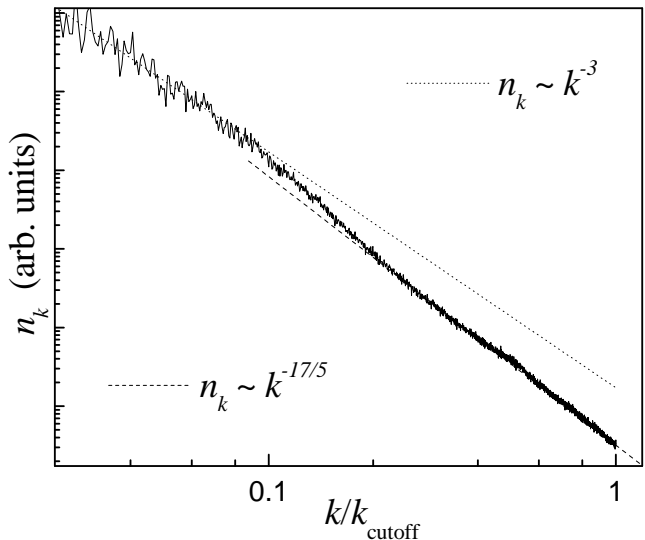}
\caption{A snapshot of time-averaged distribution of Kelvon
occupation numbers. For time averaging we use time interval which
is considerably shorter (by a factor of $\sim 3$) than the typical
time of evolution. } \label{fig_result}
\end{figure}

The key point then is to properly choose the spacing of the mesh
in accordance with the estimate (\ref{est}):
\begin{equation}
t^{(n)}_{j+1} - t^{(n)}_{j} \propto \gamma \lambda_n^2  \; ,
\label{spac}
\end{equation}
where $\gamma$ is a global ($n$-independent) small parameter
controlling the accuracy of the scheme. (The scheme is
asymptotically exact in the limit of $\gamma \to 0$.) With such a
scaling of $t^{(n)}_{j+1} - t^{(n)}_{j}$ the long-range
interactions do not form a bottleneck for the efficiency of the
numeric process. Indeed, to numerically evaluate the integral
(\ref{int_n}) one has to perform $\propto \lambda_n$ operations,
but the time interval between two integrations scales like
$\propto \lambda_n^2$, which means that on average this
integration costs only $\propto 1/\lambda_n$ operations per unit
time. To control the accuracy, we perform a test simulation (at a
smaller system size) and compare the results to the simulation of
full BSE. At a reasonably small $\gamma$, we see no difference (up
to a statistical noise, like the one noticeable in
Fig.~\ref{fig_result}) in the evolution of the distribution of
harmonics.

{\it Kelvin wave cascade}.---The decay of superfluid turbulence at
$T\to 0$ leads to a cascade of Kelvin waves generated by vortex
line reconnections \cite{Sv_95, Vinen2000}.  As it was argued by
one of us \cite{Sv_95}, in the limit of $R/a_0 \to \infty$, when
the dynamics at length scales $\sim R$ is well captured by LIA, in
the low-energy part of the inertial range there takes place a
specific regime when the energy flux in the wavenumber space is
assisted by local self-crossings of fractionalized vortex lines.
Corresponding spectrum of Kelvin waves (Kelvons) has the form
\begin{equation}
n_k \propto k^{-3}/(\ln k)^\nu  \; , \label{spec1}
\end{equation}
where $n_k$ is the Kelvon occupation number (see below) and $\nu$
is some dimensionless constant of order unity. The special role of
local self-crossings follows from the integrability of
Eq.~(\ref{LIA}) (the friction is absent at $T=0$).---The existence
of extra constants of motion excludes the possibility of the
purely non-linear cascade. Vortex line self-crossings lift these
constraints and push the cascade down to smaller wavelengths.

Vinen conjectured \cite{Vinen2000} that---apart from the
logarithmic denominator---the spectrum (\ref{spec1}) can be
derived just on the basis of dimensional considerations and is
thus insensitive to the particular cascade scenario. If true, this
statement would be of crucial importance, since at large enough
wavenumbers the cascade should radically change its microscopic
character. From dimensional estimates it is readily seen that with
increasing wavenumber non-local interactions eventually become
dominant, and the cascade is driven by purely non-linear dynamics
of Kelvons. According to Vinen, the change of the regime should
not affect the Kelvon spectrum. Recent numeric simulations
\cite{Kivotides,Vinen_2003} seem to support this idea. However, an
inconsistency can be seen in a dimensional analysis. Indeed, the
spectrum $n_k \propto k^{-3}$ implies that the typical amplitude
$b_\lambda$ of the Kelvon turbulence with the wavelength $\lambda$
is of the order of $\lambda$. This leads to a strong non-linearity
of BSE, which allows one to estimate the {\it kinetic} time
$\tau_{\rm kin}$ as the {\it dynamic} time corresponding to
$b_\lambda\sim \lambda$. It follows then that $\tau_{\rm
kin}(\lambda) \propto \lambda^2$. With this estimate for the
kinetic time, the spectrum $n_k \propto k^{-3}$ is inconsistent
with the $\lambda$-independent energy flux (see, e.g.,
Ref.~\cite{SvK_2004} for the relation between spectrum, kinetic
time, and energy flux). In our recent paper \cite{SvK_2004}, we
argue that the regime of the pure Kelvin wave cascade is
characterized by the inequality
\begin{equation}
\alpha_\lambda \equiv \, {b_\lambda \over \lambda}\, \ll \, 1  \;
, \label{non}
\end{equation}
with the parameter $\alpha_\lambda$ becoming progressively smaller
with decreasing $\lambda$.  This small parameter allows one to
develop a quantitative kinetic theory that predicts the spectrum
\begin{equation}
n_k \propto k^{-17/5} \; . \label{spec2}
\end{equation}
In terms of $b_{\lambda}$, Eq.~(\ref{spec2}) is equivalent to
\begin{equation}
b_{\lambda} \propto  \lambda ^ {6/5} \; , \label{KW amplitude
pure}
\end{equation}
while the Vinen's spectrum means $b_{\lambda} \propto \lambda$,
that is $\alpha_\lambda \sim 1$ for any $\lambda$.

Since the number $17/5=3.4$ is rather close to $3$, the accuracy
of the data of Refs.~\cite{Kivotides,Vinen_2003} turns out to be
insufficient to distinguish between the two exponents.

{\it Numeric simulation}.---The main goal of our simulation is to
observe the exponent $17/5$, and thus to verify the theory of
Ref.~\cite{SvK_2004}.

We consider a straight vortex line  along the Cartesian
$z$-direction with small ($\alpha_\lambda \ll 1$) distortions in
the $(x,y)$-plane; with the periodic boundary conditions. We use
the $z$-coordinate as a parameter of the curve, and introduce the
complex-valued function $w(z,t)=x(z,t)+iy(z,t)$ to specify the
configuration of the vortex line at a given time moment $t$. To
proceed numerically, we employ the Hamiltonian approach in terms
of the discrete complex canonical variable
$w_j(t)=x_j(t)+iy_j(t)$, $j=1,2,3,\ldots , N$, with the
Hamiltonian function [mesh spacing is set equal to unity]
\begin{eqnarray}
  \nonumber H = C \sum_{j=1}^N \sqrt{1 + |w_{j+1} - w_j|^2}+ \\
 \!  \sum_{j,l=1}^N \!\!{1 \! +\! [(w_{j+1} - w_{j-1})(w^*_{l+1} - w^*_{l-1})/8 +
{\rm c.c.}] \over \sqrt{(j-l)^2 + |w_{j}- w_{l}|^2}} \, ,
\label{H}
\end{eqnarray}
and corresponding equation of motion
\begin{equation}
i\dot{w}_j = \frac{\partial H}{\partial w_j^*} \; .   \label{EM}
\end{equation}
Here $C$ is a constant of order unity. In the long-wave limit, the
discreteness is irrelevant and the model (\ref{H})-(\ref{EM}) is
equivalent to the Hamiltonian form of BSE in terms of the function
$w(z,t)$, with $z\equiv j$ \cite{Sv_95}. From the theoretical
point of view, the particular value of $C$, and the very presence
of the first term in (\ref{H}), is not important. In practice, we
introduce this local term to render the model well-behaved at
large momenta, where the spectrum of Kelvons is very sensitive to
the short-range structure of the Hamiltonian. In our simulations,
we used $C = 23$.

The Kelvon occupation number is given by the square of the
absolute value of the Fourier transform of $w_j$ \cite{SvK_2004}:
\begin{equation}
n_k=|w_k|^2 \; ,   \label{occ}
\end{equation}
\begin{equation}
w_k\, =\, N^{-1/2}\sum_{j=1}^N\,  w_j \, {\rm e}^{ikj} \; ,
\label{F}
\end{equation}
where $k=2\pi m/N$, $m\in [-(N-1)/2 \, , \; (N-1)/2]$ is an
integer (we used odd $N$).

There are different ways to generate a Kolmogorov-type cascade.
Perhaps the most natural one is to create a steady-state regime
with a source and drain, as it was done in the simulation of
Ref.~\cite{Vinen_2003}. However, an external source (random
low-frequency force) is not necessary, since the cascade, by its
definition, is a quasi-steady-state phenomenon, where the behavior
of  short-wavelength harmonics is adjusted to the energy flux
coming from the long-wave harmonics. In a quasi-steady-state
regime, the role of a source can be played by just the
long-wavelength harmonics, as, e.g., in the simulation of
Ref.~\cite{Kivotides}. In both cases---with or without an external
force---the cascade regime requires some time to develop. We
employ the following trick to circumvent the problem of transient
period. We start with occupation numbers that progressively exceed
the quasi-steady-state ones in the direction of shorter
wavelengths. With such an initial condition, the cascade develops
in the form of a wave in the momentum space propagating from the
region of higher wavenumbers towards the region of lower
wavenumbers, the distribution {\it behind} the wave corresponding
to the quasi-steady-state cascade \cite{Sv91}. For our purpose of
resolving the spectrum (\ref{spec2}) from the spectrum $n_k
\propto k^{-3}$, it is convenient to use the latter distribution
of occupation numbers as an initial condition.

Introducing a drain---a decay mechanism for high-frequency
harmonics---significantly improves the quality of simulation,
because the large-amplitude high-frequency modes distort the
short-wavelength part of the cascade spectrum. We arrange the
drain  by periodically eliminating all the harmonics with $|k|>
k_{\rm cutoff}$, with $k_{\rm cutoff} = \pi /5$. Note that the
momentum $k_{\rm cutoff}$ is $5$ times smaller than the largest
momentum in our system. We find it important to severely cut off
the high-frequency harmonics because their dispersion strongly
differs from the parabolic dispersion of Kelvin waves in the
continuous space.

We simulated a system of the size $N=16385$. Our main results are
presented in Fig. \ref{fig_result}. We do not show a large
low-frequency part of the distribution with the spectrum $n_k
\propto k^{-3}$. Having these ballast long-wave harmonics is
important to guarantee that the inertial range is far away from
the region where the wavelength is comparable to the system size
and the kinetics is strongly affected by the wavenumber
discreteness.

Fig. \ref{fig_result} reveals the wave converting the initial
distribution $n_k \propto k^{-3}$ into a new power-law
distribution $n_k \propto k^{-\beta}$. Quantitative analysis shows
that the exponent $\beta $ coincides with $17/5$ with an absolute
error less than $0.1$, in an excellent agreement with the
prediction of Ref.~\cite{SvK_2004}.

Unfortunately, we were unable to check the prediction of
Ref.~\cite{SvK_2004} for the relation between the energy flux and
the amplitude $A$ of the distribution $n_k = A k^{-17/5}$. The
problem is in calculating the mean value of the energy flux. With
our quasi-steady-state setup, we can use only time
averaging---with an essentially limited averaging period. This
turns out to be insufficient to suppress huge fluctuations of the
instant values of the energy flux. For these purposes, the
steady-state setup of Ref.~\cite{Vinen_2003} seems to be more
adequate.

In conclusion, we have proposed an efficient numeric approach for
simulating the dynamics of superfluid turbulence in the framework
of Biot-Savart equation. The general principle is to use a less
detailed addressing for longer-range interactions. The idea is
implemented in a scheme for simulating Kelvin wave cascade on a
single vortex line. With the new scheme we were able to accurately
resolve the spectrum of the pure Kelvin wave cascade, observing an
excellent agreement with the prediction of the kinetic theory of
Ref.~\cite{SvK_2004}.

The research was supported in part by the National Science
Foundation under Grant No. PHY99-07949.

\end{document}